\documentclass[letterpaper, 10pt]{article}

\begin{document}

\twocolumn[
\begin{center}
{\Large\bf Understanding BCS Theory}\\
\vspace{0.15in}

X. H. Zheng\\
\vspace{0.05in}
\small
\em Department of Pure and Applied Physics, The Queen's University of Belfast,
Belfast BT7 1NN, Northern Ireland \em\\
\end{center}

\small
\begin{quote}
New calculation reveals that $E$ is constant in a thin layer across 
the Fermi surface, befitting the definition of energy gap parameter, 
$\Delta$ varies dramatically.  The BCS self-consistent equation has 
a simple and exact solution, showing that the well-known ratio 3.5 
must be replaced by 4.  A simple formula is found to estimate energy 
gap from first principles with reasonable accuracy, useful to the 
current research into high $T_c$ superconductivity.\\

\end{quote}]

According to Bardeen, Schrieffer and Cooper (BCS) the 
superconducting energy gap is measured by the minimum value of $E$.
In the original approximation $\Delta$ is a constant that equals
the minimum value of $E$~\cite{BCS}.  In the literature $2\Delta$ 
has become synonymous with energy gap.  However, we find that $E$ 
is constant in a thin layer across the Fermi surface, befitting the 
definition of energy gap parameter, $\Delta$ varies dramatically.  
Since $\Delta$ has played such a pivoting role in the current study 
of superconductivity, this matter appears to be serious enough to 
warrant attention.

While the BCS theory is deep and fundamental, specific results from 
this theory could be improved, because these results arise from 
approximations employed to ease mathematical difficulties.  In 
particular the BCS self-consistent equation at $T > 0$ has a simple 
and exact solution which can be proved by simple inspection.  This 
exact solution leads to a rather surprising conclusion that the 
well-known BCS ratio 3.5 $must$ be replaced by 4. 

Furthermore, we believe that the BCS theory has significant power of
prediction which was unnoticed previously.  We use the well-established
method of iteration~\cite{Porter} to solve the BCS self-consistent
equation.  We use the Mott-Jones assumption as our first approximation
\cite{Mott} which leads to a simple formula allowing us to estimate energy 
gap quantitatively from first principles, the first time to our knowledge.  
We find from 14 superconducting metals that on average the energy gap is 
$15.7\times 10^{-4}$eV, compared with the experimental value $11.3 
\times 10^{-4}$eV.  This could be interesting, in the light of the current 
intensive effort to search how high $T_c$ might be for various novel 
superconductive materials.

We start with the BCS self-consistent equation at $T > 0$:
\begin{equation}
\Delta({\bf k})=\sum_{\bf q}V({\bf k, q})\frac{\Delta({\bf k + q})}
{E({\bf k + q})}\tanh\left[\frac{E({\bf k + q})}{2k_BT}\right]
\label{eq:self-consistent(T>0)}
\end{equation}
where $E = (\Delta^2 + \epsilon^2)^{1/2}$, $\epsilon$ is the electron
energy relative to the Fermi level, {\bf k} and {\bf q} electron and 
phonon vectors, respectively~\cite{BCS}.  We solve this equation 
through iteration~\cite{Porter}.  As our first approximation, we
assume $\epsilon({\bf k + q})\approx\epsilon({\bf k})$ which was 
used successfully by Mott and Jones to study metal conductivity 
for the reason that, when colliding with atoms, electrons tend to 
change momentum rather than jump into orbits already occupied by 
other electrons \cite{Mott}.  In addition to this, we are going 
to see that $\epsilon({\bf k + q})\approx\epsilon({\bf k})$ is 
actually the condition to have a superconducting gap.  We consider 
a spherical Fermi surface, so that both $\Delta$ and $E$ depend only 
on $\epsilon$.  Therefore our assumption $\epsilon({\bf k + q})
\approx\epsilon({\bf k})$ ensures $\Delta({\bf k + q})\approx
\Delta({\bf k})$ and $E({\bf k + q})\approx E({\bf k})$, which 
lead through Eq.~(\ref{eq:self-consistent(T>0)}) to
\begin{equation}
E = E_0\tanh\left(\frac{E}{2k_BT}\right)
\label{eq:gap(T>0)}
\end{equation}
where
\begin{equation}
E_0 = \sum _{\bf q}V({\bf k,q})
\label{eq:gap}
\end{equation}
which is just the value of $E$ at $T = 0$: $\tanh(E/2k_BT)
\rightarrow 1$ when $T\rightarrow 0$ in Eq.~(\ref{eq:gap(T>0)}).

The iteration method was not used by BCS to solve 
Eq.~(\ref{eq:self-consistent(T>0)}).  Instead, they assumed $V = const.$ 
in that equation, and converted the summation there into integration,
giving
\begin{equation}
\frac{1}{N(0)V} = \int^{\hbar\omega}_{0}\tanh\left(\frac{E}{2k_BT}
\right)\frac{d\epsilon}{E}
\label{eq:self-consistent(BCS)}
\end{equation}
which is one of the key equations of the BCS theory, where $N(0)$ is the 
density of state near the Fermi surface, $\omega$ an unspecified phonon
frequency.  Further approximation was applied to find the well-known 
ratio $2\Delta /k_BT_c = 3.5$~\cite{BCS}.  However, it is interesting 
that $E$ from Eq.~(\ref{eq:gap(T>0)}) is also the $exact$ solution of 
Eq.~(\ref{eq:self-consistent(BCS)}), provided that $\hbar\omega N(0)V = E_0$:
approximation is not necessary.  This can be checked easily by simple 
inspection.  Since $\tanh(E/2k_BT)\leq E/2k_BT$, we find from 
Eq.~(\ref{eq:gap(T>0)}) $T\leq E_0/2k_B$ which means $T_c = E_0/2k_B$ 
and therefore 
\begin{equation}
\frac{2E_0}{k_BT_c} = 4
\label{eq:critical-ratio}
\end{equation}
which reduces to $2\Delta/k_BT_c = 4$ when $\epsilon = 0$.  Therefore the
ratio 3.5 $must$ be replaced by 4, regardless of whether 
Eq.~(\ref{eq:self-consistent(T>0)}) is solved via the original BCS 
approximation or not.  Indeed, the ratio 3.5 does not appear to have 
much advantage: we find from 23 superconducting elements in the literature 
\cite{Poole} that on average $2E_0/k_BT_c = 3.8$ with standard deviation 
0.597.  The deviation is 0.627 from the ratio 4.0, compared with 0.672
from ratio 3.5.

Now we evaluate $E_0$ in Eq.~(\ref{eq:gap}) quantitively.  We have 
\begin{equation}
V({\bf k , q} )=\sum_{l=1}^{3}\frac{2\hbar\omega_{l}({\bf q})
\mathcal{M}_{l}^2({\bf k,q})}{[\hbar\omega_{l}({\bf q})]^{2}-
[\epsilon({\bf k + q})-\epsilon({\bf k})]^{2}}
\label{eq:c-number}
\end{equation}
where $\omega _{l}$ is the phonon frequency, $l$ identifies phonon 
branch (excluding transverse phonons, which do not interact with 
electrons in $N$-processes) \cite{Ziman}.  Eq.~(\ref{eq:c-number})
is slightly different from that in \cite{BCS} where phonon
polarization is neglected.  The matrix element 
\begin{equation}
\mathcal{M}_l({\bf k, q})= \tilde{q}_l\!\left[\frac{\hbar N}
{2M\omega_l({\bf q})}\right]^{1/2}\!\!\!\int_{\Omega}\psi_{{\bf k+q},
\sigma}^{*}({\bf r})\delta \mathcal{V}({\bf r})\psi_{{\bf k},\sigma}
({\bf r})d{\bf r}
\label{eq:matrix}
\end{equation}
measures the strength of electron-phonon interaction.  Here 
$\psi$ is the electron wave function, $\sigma$ spin $\uparrow$ 
or $\downarrow$, $M$ mass of an atom, $N$ number of atoms in unit 
volume, {\bf r} coordinates in real space, $\Omega_{0}$ a volume 
surrounding the atom, $\Gamma_{0}$ its boundary, and $\delta\mathcal{V}
({\bf r}) = \mathcal{V}({\bf r}) - \mathcal{V}(\Gamma_{0})$, 
$\mathcal{V}$ being the potential field.  We define $\tilde{q}_l$ 
as the $l$-th component of $U{\bf q}$, $U$ being the $3 \times 3$ 
unitary matrix found when solving the classical equation of motion 
for the atom.  Mott and Jones found matrix elements when $\Omega_{0}$ 
is the Wigner-Seitz cell~\cite{Mott}.  We find Eq.~(\ref{eq:matrix}) 
when $\mathcal{V}(\Gamma_{0})$ is constant (this defines $\Omega_{0}$ 
in a natural manner). 

We use free electron energy to evaluate Eq.~(\ref{eq:c-number}). 
This means that the Fermi surface is spherical, our assumption 
when solving Eq.~(\ref{eq:self-consistent(T>0)}).  As a result, 
the denominator of Eq.~(\ref{eq:c-number}) becomes $4\epsilon 
_{F}\epsilon _{\bf q}[\delta _{l}^{2} - (\zeta + \cos\theta)^{2}]$, 
$\epsilon _{F} = (\hbar^{2}/2m)|{\bf k}|^{2}$ is the Fermi energy 
(we study $|{\bf k}|$ near the Fermi surface), $\epsilon _{\bf q} 
= (\hbar^{2}/2m)|{\bf q}|^{2}$, $\delta _{l}^{2} = (m/2)v_{l}^{2}/
\epsilon _{F}$, $v_{l} = \omega _{l}({\bf q})/{\bf q}|$ the sound 
velocity.  In the Debye approximation $\delta _{l} = (Z/16)^{1/3}
\Theta _{D}/\Theta _{F}\approx 10^{-3}$ in all superconducting 
metals, where $Z$ is the valency, $\Theta _{D}$ and $\Theta _{F}$ 
are the Debye and Fermi temperatures.  It is apparent that 
$V({\bf k, q}) > 0$ (condition to have an energy gap) holds in 
Eq.~(\ref{eq:c-number}) only when $\zeta + \cos\theta \approx 0$, 
$\zeta = |{\bf q}|/|2{\bf k}|$, $\theta$ being the angle between 
{\bf k} and {\bf q}, so that $|{\bf k + q}|^{2} = |{\bf k}|^{2} 
+ |{\bf q}|^{2} + 2|{\bf k}||{\bf q}|\cos\theta\approx|{\bf k}|^{2}$.  
Thus $\epsilon({\bf k + q})\approx\epsilon({\bf k})$: electrons 
change momentum but not energy in scattering, the assumption by 
Mott and Jones~\cite{Mott}.

We substitute Eq.~(\ref{eq:c-number}) into Eq.~(\ref{eq:gap})
and replace the summation over {\bf q}  with an integration over 
$(4\pi/3)k_{D}^{3}$ (volume of the first Brillouin zone), which 
exists in the sense of the Cauchy principal value (used by Kuper 
to verify the BCS theory) \cite{Porter, Kuper}, i.e.\ positive and 
negative contributions of $V({\bf k, q})$, if finite, are cancelled 
on a series of spherical surface, the singular point ignored.  We 
are entitled to do so, because Eq.~(\ref{eq:c-number}) is defined 
on a grid of {\bf k} and {\bf q}, which may not be in precise 
combinations to let $V({\bf k, q}) = \infty$.  We can also avoid 
such combinations by suppressing a few phonons with little physical 
consequence.  This principal value varies little among phonon branches, 
allowing us to use $\sum _{l}\tilde{q}_{l}^{2} = {\bf q}^{t}U^{t}
U{\bf q} = |{\bf q}|^{2}$ ($U$ unitary) to simplify the result.  
We use the expression for metal resistivity \cite{Mott} to calibrate 
$\delta\mathcal{V}$ in Eq.~(\ref{eq:matrix}) and find
\begin{equation}
E_0 = \frac{\hbar e^2}{k_BT_{\rho}}\eta n \rho v^2
\label{eq:gap-evaluated}
\end{equation}
where $e$ and $n$ are electron charge and density, respectively, $k_B$
the Boltzmann constant, $\rho$ the resistivity at temperature $T_\rho$ 
(not necessarily low), $v = k_B\Theta _D/\hbar k_D$ the Debye sound
velocity, $k_D$ being the phonon cut-off wavenumber, and
\begin{equation}
\eta = \frac{1}{\pi}\int_{0}^{(4Z)^{-1/3}}\!\!\!\!F^{2}(x)
\frac{\zeta^{2}d\zeta}{1 - \zeta^{2}}\bigg/
\int_{0}^{(4Z)^{-1/3}}\!\!\!\!F^{2}(x)\zeta^{3}d \zeta 
\approx 1
\label{eq:eta}
\end{equation}
Here $F(x) = 3(x \cos x - \sin x)/x^{3}$ is the overlap integral function, 
$x = 3.84\alpha^{1/3}Z^{1/3}\zeta$, $\alpha = N \Omega _{0}/\Omega$ the 
fraction of $\Omega _{0}$ in a primitive cell, and $\Omega$ the unit 
volume.  We assume ${|\bf q}|/|2{\bf k}| = \zeta <(4Z)^{-1/3} < 1$, 
because Eq.~(\ref{eq:c-number}) arises from a canonical transformation
\cite{Frohlich} where operator commutation requires 
${\bf q}\neq\pm 2{\bf k}$.\\

\begin{picture}(230,120)(0,0)
\put(-11,0){\framebox(230,120)}

\put(0,100){(A)}
\thinlines
\put(0,30){\line(1,0){90}}
\put(90,30){\line(-5,1){10}}
\put(90,30){\line(-5,-1){10}}

\put(60,30){\line(0,1){70}}
\put(60,100){\line(1,-5){2}}
\put(60,100){\line(-1,-5){2}}

\put(65,35){$0$}
\put(90,35){$\epsilon$}
\put(40,95){$E_{\bf k}$}

\put(40,20){\line(0,-1){10}}
\put(80,20){\line(0,-1){10}}
\put(40,15){\line(1,0){7}}
\put(80,15){\line(-1,0){7}}
\put(55,12){$2E$}

\thicklines
\put(00,90){\line(1,-1){40}}
\put(40,50){\line(1,0){40}}
\multiput(40,50)(2,-2){10}{\circle*{1}}
\put(65,55){$E$}

\put(110,100){(B)}
\thinlines
\put(110,30){\line(1,0){90}}
\put(200,30){\line(-5,1){10}}
\put(200,30){\line(-5,-1){10}}

\put(155,30){\line(0,1){70}}
\put(155,100){\line(1,-5){2}}
\put(155,100){\line(-1,-5){2}}

\put(160,35){$0$}
\put(200,35){$\epsilon$}
\put(139,95){$\Delta$}

\put(135,20){\line(0,-1){10}}
\put(175,20){\line(0,-1){10}}
\put(135,15){\line(1,0){7}}
\put(175,15){\line(-1,0){7}}
\put(150,12){$2E$}

\thicklines
\qbezier(135,30)(155,70)(175,30)
\put(160,55){$E$}
\end{picture}\\

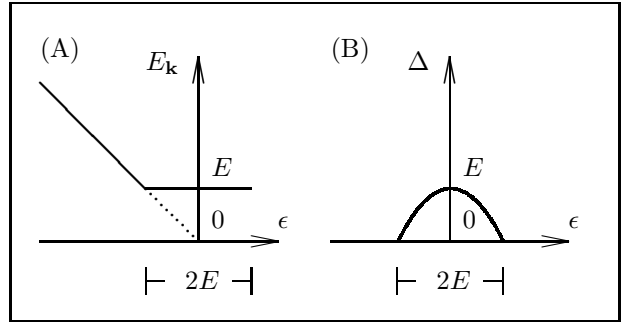
\noindent FIG.~1.\ (A) Extended solution of the BCS self-consistent 
equation, $E_{\bf k} = E$ or $-\epsilon$, when $-E < \epsilon < E$ or 
$\epsilon < -E$, respectively.  (B) $\Delta = (E^2 - \epsilon^2)^{1/2}$ 
which varies dramatically when $-E < \epsilon < E$.  $\Delta\approx E$ 
only when $\epsilon\approx 0$.\\

It is apparent from Eqs.~(\ref{eq:gap(T>0)}) and (\ref{eq:gap-evaluated})
that $E$ and $E_0$ are constant.  In contrast $\Delta = (E^2 - \epsilon^2)
^{1/2}$ varies dramatically when $-E < \epsilon < E$.  Beyond this range 
Eq.~(\ref{eq:self-consistent(T>0)}) does not have non-trivial solution 
in first iteration.  This justifies the approach of BCS to integrate 
Eq.~(\ref{eq:self-consistent(T>0)}) only in a thin layer across the Fermi 
surface \cite{BCS}.  However Eq.~(\ref{eq:self-consistent(T>0)}) has an
extended solution (Fig.~1)
\begin{equation}
E_{\bf k} = \left\{\begin{array}{r@{\quad:\quad}r}E & -E < \epsilon < E\\
-\epsilon & \epsilon < -E \end{array}\right.
\label{eq:dispersion}
\end{equation}
where $E_{\bf k} = -\epsilon$ arises from the trivial solution $\Delta = 0$ 
of Eq.~(\ref{eq:self-consistent(T>0)}).  The pair occupancy (Fig.2)
\begin{equation}
h_{\bf k} = \frac{1}{2}\left(1 - \frac{\epsilon}{E_{\bf k}}\right)
\label{eq:pair-occupancy}
\end{equation}
varies linearly when $-E < \epsilon < E$, but equals 1 when $\epsilon < -E$.
This also justifies the BCS approach to integrate 
Eq.~(\ref{eq:self-consistent(T>0)}) only in a thin layer: otherwise we have 
$h_{\bf k} > 1$ or $< 0$.  The over-all probability of occupancy at $T > 0$ 
is given by $h_{\bf k}(1 - 2f_{\bf k})$.  Here
\begin{equation}
f_{\bf k} = 1\Bigg/\left[1 + \exp\left(\frac{E_{\bf k}}{k_BT}\right)\right]
\label{over-all-occupancy}
\end{equation}
is the probability of excitation which, unlike the Fermi-Dirac distribution, 
drops towards the interior of the Fermi sea (Fig.~2) where 
electrons are apparently more difficult to excite.  Curves in Figs.\ 1 and 2
all have a kink at $\epsilon = -E$, which is likely to be smoothed out after 
further iterations.\\

\begin{picture}(230,120)(0,0)
\put(-11,0){\framebox(230,120)}

\put(0,100){(A)}
\thinlines
\put(0,30){\line(1,0){90}}
\put(90,30){\line(-5,1){10}}
\put(90,30){\line(-5,-1){10}}

\put(45,30){\line(0,1){70}}
\put(45,100){\line(1,-5){2}}
\put(45,100){\line(-1,-5){2}}

\put(35,35){$0$}
\put(85,35){$\epsilon$}
\put(25,95){$h_{\bf k}$}

\put(25,20){\line(0,-1){10}}
\put(65,20){\line(0,-1){10}}
\put(25,15){\line(1,0){7}}
\put(65,15){\line(-1,0){7}}
\put(40,12){$2E$}

\put(40,80){\line(1,0){10}}
\put(55,80){$1$}

\thicklines
\put(0,80){\line(1,0){25}}
\put(25,80){\line(4,-5){40}}

\put(110,100){(B)}
\thinlines
\put(110,30){\line(1,0){90}}
\put(200,30){\line(-5,1){10}}
\put(200,30){\line(-5,-1){10}}

\put(165,30){\line(0,1){70}}
\put(165,100){\line(1,-5){2}}
\put(165,100){\line(-1,-5){2}}

\put(155,35){$0$}
\put(200,35){$\epsilon$}
\put(145,95){$f_{\bf k}$}

\put(145,20){\line(0,-1){10}}
\put(185,20){\line(0,-1){10}}
\put(145,15){\line(1,0){7}}
\put(185,15){\line(-1,0){7}}
\put(160,12){$2E$}

\put(160,55){\line(1,0){10}}
\put(175,63){1/2}

\thicklines
\put(145,50){\line(1,0){40}}
\qbezier(110,35)(133,37)(145,50)
\end{picture}\\

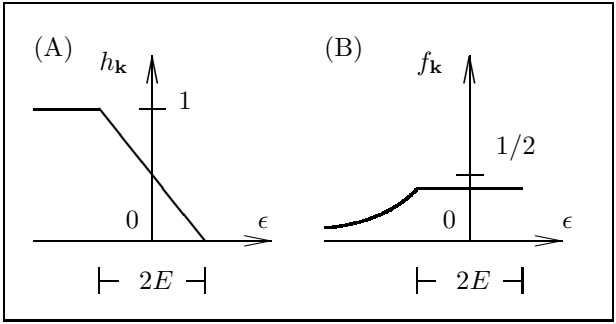
\noindent FIG.~2.\ (A) Electron pair occupancy, and (B) schematic of 
excitation probability.\\

In the original BCS paper $E_{\bf k} = (\Delta^2 + \epsilon^2)^{1/2}$ 
is known as the dispersion law~\cite{BCS}.  In Eq.~(\ref{eq:dispersion}) this 
dispersion law is reduced to the normal law once $\epsilon < -E$, consistent 
with the physics that no electron-phonon interaction takes place deep inside 
the Fermi sea.  In the view of BCS the energy gap is measured by the minimum
value of $E_{\bf k}$~\cite{BCS}.  It is clear from Fig.~1
that $E$ fits this definition.  In comparison $\Delta$ appears to be a 
parameter to measure the electron-phonon interaction, which weakens 
dramatically with increasing distance from the Fermi surface, and cuts 
off when $|\epsilon| > E$.  Furthermore, $E_0$ in Eq.~(\ref{eq:gap-evaluated})
matches experimental values of the superconducting gaps of metals reasonably well, 
despite that on surface the combination of various fundamental constants in 
that equation appears to be a very long shot.  Apparently, a good superconductor 
must have numerous free electrons (large $n$) scattered frequently by atoms 
(large $\rho$) moving quickly to facilitate pairing (large $v$).  When $\alpha 
= 1$, Eq.~(\ref{eq:gap-evaluated}) yields $2E_0 = 2.2$, 18 and 27 for Cd, Ta 
and Nb (1.5, 14 and 30.5 experimentally, in $10^{-4}$eV), which are of the 
right order, although over and under-estimations are possible.  On average 
Eq.~(\ref{eq:gap}) yields $2E_0 = 15.7$ for Zn, Cd, Hg, Al, Ga, Tl, Sn, Pb, 
V, Nb, Ta and Mo (11.3 experimentally).

Although accepted by many, $\Delta = const.$ is from approximation, employed 
to ease the mathematical difficulty to evaluate the BCS theory, and therefore
not as fundamental as the theory itself.  In fact $\Delta = const.$ leads to
several conclusions that have to be justified.  For example $E = (\Delta^2
+ \epsilon^2)^{1/2}$ increases when $\epsilon > 0$.  It is not clear why pairs
above the Fermi surface are more difficult to break, despite their higher
energy.  The dispersion law $E_{\bf k} = (\Delta^2 + \epsilon^2)^{1/2}$ 
reduces to the normal law only when $|\epsilon| >> \Delta$.  It is not clear
why this normal law is distorted inside the Fermi sea, where no 
electron-phonon interaction is supposed to take place.  Furthermore
$\Delta = const.$ leads to Eq.~(\ref{eq:self-consistent(BCS)}) which is not 
easy to be compared with experimental data.

In conclusion the BCS theory has significant power of prediction when
evaluated via the well-established method of iteration.  This could 
be of current interest, because some novel theories for high $T_c$ 
superconductors resemble the BCS theory quite closely~\cite{Waldram,
Anderson}.  These theories are often phenomenological, but nonetheless
detailed enough for one to evaluate the matrix element~\cite{Anderson}.  
Therefore a formula similar to Eq.~(\ref{eq:gap-evaluated}) might arise,
which could be useful when comparing the theory with experiments.


\end{document}